\documentclass[twocolumn,superscriptaddress,aps,prb,amsmath,amssymb]{revtex4-1}
\usepackage{graphicx}
\usepackage{dcolumn}
\usepackage{bm}
\usepackage{epsfig}
\usepackage[usenames]{color}
\usepackage{amsthm} 
\usepackage[colorlinks=true,linkcolor=blue,citecolor=blue,urlcolor=blue]{hyperref}
\usepackage{setspace}
\usepackage{color}

\let\baraccent=\=

\renewcommand{\v}[1]{\ensuremath{\mathbf{#1}}}

\renewcommand{\=}[1]{\stackrel{#1}{=}}  
 
\newcommand{\unit}[1]{\ensuremath{\hat{#1}}} 
\newcommand{\uv}[1]{\ensuremath{\mathbf{\hat{#1}}}}

\newcommand{\braket}[2]{\left< #1 \vphantom{#2} \right| \left. #2 \vphantom{#1} \right>} 
\newcommand{\matrixel}[3]{\left< #1 \vphantom{#2#3} \right| #2 \left| #3 \vphantom{#1#2} \right>}

\newcommand{\al}[1]{\begin{align}#1\end{align}}
\newcommand{\bs}{\begin{split}}
\newcommand{\es}{\end{split}}

\newcommand{\m}[1]{$ #1 $}
\newcommand{\kp}{$\v k \cdot \v p~$}

\def\kv{{\bf k}}
\def\rv{{\bf r}}
\def\Av{{\bf A}}
\def\be{\begin{equation}}
\def\ee{\end{equation}}

\begin{document}
\title{Derivation of effective spin-orbit Hamiltonians and spin lifetimes, with application to SrTiO$_3$ heterostructures}
\author{C\"{u}neyt \c{S}ahin}
 \email{cuneyt-sahin@uiowa.edu}
\affiliation{Optical Science and Technology Center and Department of Physics and Astronomy, University of Iowa, Iowa City, Iowa 52242, USA}
\author{Giovanni Vignale}
\affiliation{Department of Physics and Astronomy, University of Missouri, Columbia, Missouri 65211, USA}
\author{Michael E. Flatt\'e}%
\affiliation{Optical Science and Technology Center and Department of Physics and Astronomy, University of Iowa, Iowa City, Iowa 52242, USA}
\date{\today}
\begin{abstract}
 A general approach is derived for constructing an effective spin-orbit Hamiltonian for nonmagnetic materials, which is useful for calculating spin-dependent properties near an arbitrary point in momentum space with pseudospin degeneracy. The formalism is verified through comparisons with other approaches for III-V semiconductors, and its general applicability is illustrated by deriving the spin-orbit interaction and predicting spin lifetimes for strained SrTiO$_3$ and a two-dimensional electron gas in SrTiO$_3$ (such as at the LaAlO$_3$/SrTiO$_3$ interface).  These results suggest robust spin coherence and spin transport properties in SrTiO$_3$-based materials at room temperature.
\end{abstract}
\pacs{pacs numbers}
\maketitle

\section{Introduction}
Spin dynamics in nonmagnetic wide-bandgap materials has received renewed attention due to the exceptionally long spin coherence times of spin centers in diamond\cite{Balasubramanian2009} and silicon carbide\cite{Koehl2011}, interest in spin injection into bulk doped SrTiO$_3$ (STO)\cite{Han2013} as well as  Rashba coefficients\cite{Caviglia2010} and spin injection\cite{Reyren2012} in the strain-tunable and growth-tunable\cite{Jalan2011,Son2010} high-density, high-mobility two-dimensional electron gas (2DEG) at the interface\cite{Ohtomo2004} between  LaAlO$_3$ and SrTiO$_3$ (LAO/STO).
 For well-explored materials such as III-V semiconductors and their heterostructures, effective pseudomagnetic fields\cite{Meier1984,Dresselhaus1955} arising from the inversion asymmetry of the crystal split most degeneracies of the electronic states, and these fields dominate spintronic properties of the material such as spin lifetimes\cite{Dyakonov1972,Lau2001}. 
 
 When materials have inversion symmetry, however, these fields vanish and the subtle spin-orbit entanglement of the wave functions controls spintronic properties\cite{Yafet1963} and dominates spin lifetimes through the scattering-driven Elliot-Yafet process\cite{Meier1984}. The construction of effective spin-orbit Hamiltonians for nonmagnetic scattering in semiconductors, that include the spin-orbit entanglement of the wave functions\cite{Li2011,Gmitra2013}, usually proceeds from a simple effective model of the material\cite{Yafet1963,Meier1984}, especially when only a small number of invariants are allowed by symmetry\cite{BirPikus}.
If such simple effective models are not apparent, such as for indirect-gap, multivalley semiconductors ({\it e.g.} diamond) or for single-valley bands with orbital degeneracy ({\it e.g.}  the $d$-character conduction band of STO), then the process to construct a spin-orbit Hamiltonian is not clear. {\it ad hoc} and specialized approaches\cite{Li2011,Gmitra2013} can miss properties apparent in a more complete tight-binding approach\cite{Tang2012} or other full-zone approach\cite{Restrepo2012}. A  formal prescription to construct an effective spin-orbit Hamiltonian is required, built off a full-zone description of the electronic structure.

Here  a rigorous prescription for the construction of such an effective spin-orbit Hamiltonian near a point of pseudospin degeneracy is provided and applied to materials that are spatially inversion symmetric with doubly degenerate bands.  We verify  this prescription  by testing it at the Brillouin zone center of direct-gap III-V semiconductors (where there is double degeneracy), comparing the Hamiltonian and spin lifetimes from a tight-binding band structure  to those from a \kp model  describing the single conduction valley. We then extract from this formalism an effective spin-orbit Hamiltonian for STO, and use it to predict spin lifetimes for conduction electrons in   strained STO and an LAO/STO 2DEG. We find exceptionally long spin lifetimes in both, suggesting that STO-based materials should have robust room-temperature spintronic properties. This prescription to construct an effective spin-orbit Hamiltonian should also be of assistance in calculating a broad assortment 
of spin-related properties, including spin diffusion lengths, spin Hall conductivities, $g$-tensors, and spin precession lengths, relying on valid electronic structure calculations from a range of approaches. Thus it provides a complement to density-functional-theory-based calculations of spin lifetimes\cite{Restrepo2012,Fedorov2013}, which assume Kohn-Sham wave functions and energies accurately represent the material's single-particle properties, and are also challenging to implement for heterostructures. 

\section{Formalism} In systems with time-reversal invariance and spatial inversion symmetry, the electronic states are (at least) doubly degenerate at each crystal momentum ${\bf k}$, and the spintronic properties will be governed by spin-orbit entanglement in the wave functions of the electronic states. We further focus on systems with exactly two-fold state degeneracy at each ${\bf k}$. The Bloch states are denoted by $\psi_{\kv,\alpha}(\rv)=e^{i\kv \cdot \rv}u_{\kv,\alpha}(\rv)$, where $\alpha=\pm 1$  is a pseudospin index that labels the two degenerate states at each $\kv$, $u$ is a periodic function of $\rv$, and  $\psi$ and $u$ are two-component spinors.  The corresponding energies are $E_{n\kv}$, independent of $\alpha$.  The two degenerate states are connected by the combination of a time-reversal and a spatial inversion operation:
\be\label{TP}
u_{\kv,-\alpha}(\rv)=i\sigma_y u^*_{\kv,\alpha}(-\rv)\,,
\ee
where $\sigma_y$ is the $y$ Pauli matrix. This model describes germanium, silicon and diamond, as well as STO where the orbital degeneracy at the conduction band minimum has been lifted due to strain or quantum confinement. For materials in which the spin and orbital degrees of freedom are strongly mixed, the pseudospin doublet described by the $u_{\kv=0,\alpha}$ remains stable; hence here we focus on the lifetime of nonequilibrium populations of pseudospin and for simplicity of language drop the prefix ``pseudo''.

Consider a spin-orbit Hamiltonian (\textit{e.g.} a tight-binding Hamiltonian with spin-orbit interaction) describing a material, and determining the wave functions and their corresponding energies in the immediate vicinity of a valley minimum  $\mathbf k_\mu$. 
 The conduction band (labeled $``c"$)  has equivalent minima at symmetry-related points $\mathbf k_1 , \mathbf k_2 ,...,\mathbf k_N$ ({\it e.g.} $N = 4$ for Ge, $N = 6$ for Si and diamond,
 $N=1$ for  STO with strain or quantum confinement). 
Describing each valley (at $\kv_\mu$) independently, we set $\kv=\kv_\mu+\tilde \kv$ and call $\psi_{n,\alpha}(\rv)= e^{i\kv_\mu \cdot\rv} u_{n,\alpha}(\rv)$ 
the wave functions at $\kv_\mu$ (here $n$ denotes a generic band index).
These $u_{n  \alpha}(\rv)$ -- a complete set of periodic functions -- form a basis to expand the periodic part of the wave function at small, {finite} $\tilde{\mathbf k}$,
\al{\label{Expansion}
u_{c \tilde{\mathbf k}\alpha}(\mathbf r, \sigma) \simeq  u_{c  \alpha}(\mathbf r, \sigma) -i\sum_{{\beta},{n}}\tilde{\mathbf k}\cdot \mathbf A_{c\alpha, n\beta}u_{n  \beta}(\mathbf r, \sigma),
} 
where
\al{\mathbf A_{c \alpha, n \beta}\equiv i \braket{u_{n \beta}}{\partial{u_{c \tilde{\mathbf k} \alpha}}/\partial{\tilde {\mathbf k}}}_{\tilde{\mathbf k} =0}=\mathbf A^* _{n \beta , c \alpha}.
}
The value of the coefficients $\mathbf A_{c \alpha , n \alpha}$ depends on an arbitrary choice of $\kv$-dependent phase factors  $e^{i\phi_{n,\alpha}(\tilde\kv)}$, by which the Bloch wave functions at $\tilde \kv \neq 0$ can be multiplied.  This arbitrariness is reduced by insisting that the periodic parts of the wave functions $u_{\tilde \kv,\alpha}$ and $u_{\tilde \kv,-\alpha}$ be related to each other by Eq.~(\ref{TP}).  When this condition is satisfied,
\hbox{$\Av_{n\alpha,n\beta}=-\Av_{n -\beta,n-\alpha}$}, which implies
\hbox{$\Av_{n\alpha,n\alpha}=-\Av_{n-\alpha,n-\alpha} {\rm (real)}$} and 
$\Av_{n\alpha,n-\alpha}=0\,$. (See Appendix~\ref{appA})
  For $\mathbf A_{n \alpha , m \beta}$ with $n\neq m$,
a standard calculation leads to 
\al{\label{connection}
 -i\mathbf A_{n\alpha, m\beta}=\frac{\matrixel{u_{m \tilde{\mathbf k} \beta}}{\nabla_{\tilde{\mathbf k}}H_{\tilde{\mathbf k}}}{u_{n \tilde{\mathbf k} \alpha}}}{E_{n\tilde{\mathbf k}}-{E_{m\tilde{\mathbf k}}}}\,.}
 Here $H_{\tilde{\mathbf k}}$ is the Hamiltonian of the periodically translationally invariant system, so $\tilde{\mathbf k}$, the crystal momentum difference from ${\mathbf k_\mu}$, is a good quantum number, and the operator ${\nabla_{\tilde{\mathbf k}}H_{\tilde{\mathbf k}}}$ is straightforward to evaluate.

\section{The Effective External Potential}

Construction of the effective spin-orbit Hamiltonian  for a  spin-independent scalar potential $V(\mathbf r)$ (slowly varying on the unit-cell scale) requires the evaluation of  matrix elements of $V(\mathbf r)$  between conduction band states $\psi_{c \tilde{\mathbf k} \alpha}= e^{i(\kv_\mu+\tilde\kv)\cdot\rv}u_{c \tilde{\mathbf k} \alpha}$,
\be
 \tilde V_{\tilde{\mathbf k}'\alpha',\tilde{\mathbf k}\alpha}=\int d\mathbf r \psi_{c\tilde{\mathbf k}'\alpha'}(\mathbf r)V(\mathbf r)\psi_{c\tilde{\mathbf k}\alpha}(\mathbf r)\,.
 \ee
Using Eq.~(\ref{Expansion}) and  assuming $V(\mathbf r)$ varies slowly, Eq.~(\ref{matrixelement}) can be integrated over any unit cell. Summing over the unit cells and using the orthogonality properties of  $u_{n \mathbf k,\alpha}(\mathbf r)$ yields (See Appendix~\ref{appB})
\al{
\bs
 \tilde V_{\tilde{\mathbf k'}\alpha',\tilde{\mathbf k}\alpha}=& \int d\mathbf r e^{i(\tilde{\mathbf k}-\tilde{\mathbf k'})\cdot \mathbf r}\Bigg[\delta_{\alpha' \alpha}V(\mathbf r) 
 +\sum_{ij} B^{ij}_{\alpha \alpha'}\nabla_i V(\mathbf r)\nabla_j\Bigg],\label{matrixelement}
\es
}
where $i$ and $j$ can be $x$, $y$, or $z$ and 
\al{B^{ij}_{\alpha' \alpha}\equiv 
\braket{\frac{\partial {u_{c\tilde{\mathbf k}\alpha'}}}{{\partial \tilde{k_i}}} }{ \frac {\partial{u_{c\tilde{\mathbf k}\alpha}}} {\partial \tilde{k_j}}}_{\tilde{\mathbf k}=0}.
}

The symmetries of $B^{ij} _{\alpha \alpha'}$ emerge when written as an operator in 
spin space:
\al{
B^{ij}_{\alpha \alpha'}=B_{ij0}\delta_{\alpha \alpha'}+\sum_k B_{ijk}[\sigma_k]_{\alpha' \alpha}\
}
where
\al{ B_{ijk}=\frac{1}{2}\sum_{\alpha \alpha'} B^{ij}_{\alpha \alpha'}[\sigma_k]_{\alpha \alpha'}\\
B_{ij0}=\frac{1}{2}\sum_{\alpha} B^{ij}_{\alpha \alpha},}
and the index $k$ can be $x$, $y$, or $z$. 
As defined  $B_{ijk}=B^*_{jik}$, and from time reversal invariance   $B_{ij0}$ is real,  thus $B_{ij0}=\lambda _{ij}=\lambda _{ji}$.
In contrast, $B_{ijk}$ is imaginary and  antisymmetric: 
\al{B_{ijk}=i\lambda _{ijk}=-i\lambda _{jik}.}
In this notation the effective potential becomes
\begin{widetext}
\al{\label{eq:effpot}
\bs
 \tilde{V}(\uv{r},\unit{\sigma}) = \Bigg[V(\uv{r}) +\sum_{ij}\lambda_{ij}\nabla_i V(\uv{r}) \nabla_j\Bigg]
 + i\sum_{ijk}\lambda_{ijk}\nabla_i V(\uv{r}) \nabla_j \unit{\sigma}_k
 \es.
}
The  tensor $\lambda_{ijk}$, which defines the effective spin-orbit interaction in the conduction band, can be expressed exactly as

\begin{equation} \label{eq:so-tensor}
\lambda_{ijk}=\frac{1}{2}Im\sum_{\alpha \alpha'}[\sigma_k]_{\alpha \alpha'} \sum_{n\neq c, \beta} \frac{\matrixel{u_{c \tilde{\mathbf k} \alpha'}}{\nabla_{\tilde{k}_i} \unit H_{\tilde{\mathbf k}}} {u_{n \tilde{\mathbf k} \beta}} \matrixel{u_{n \tilde{\mathbf k} \beta}}{\nabla_{\tilde{k}_j} \unit H_{\tilde{\mathbf k}}}{u_{c \tilde{\mathbf k} \alpha}}}{(E_{c \tilde{\mathbf k}} -E_{n \tilde{\mathbf k}})^2} \Bigg |_{\tilde{\mathbf k}=0}\,, 
\end{equation}
\end{widetext}
where the sum runs over all the bands ($n$) other than the conduction band  ($c$) -- all intra-band contributions vanish by virtue of the identities
above.  
Eq.~(\ref{eq:so-tensor}) is independent of any arbitrary $\kv$-dependent phase factors by which the periodic parts of the Bloch wave functions may be multiplied.  This formula, a principal result of this Letter, is suitable for numerical evaluation of the effective spin-orbit interaction, provided a calculation of the periodic parts of the Bloch wave functions $u_{n \kv,\alpha}$ is available.

\section{Scattering in the effective spin-dependent potential}
Knowledge of the $\lambda$'s allows us to construct the effective spin-orbit interaction between electrons in a specific  band and scattering from a scalar spin-independent potential $V(\rv)$ ({\it e.g.} impurity scattering or phonon scattering in a quasi-elastic approximation\cite{Yu3ed}) using Eq.~(\ref{eq:effpot}). For multiple conduction bands located near a single minimum, such as for strontium titanate based materials, the individual bands have Bloch functions that are orthogonal to each other at the conduction minimum, so scattering between bands is inefficient. In contrast the largest contribution to scattering will come from scattering within specific bands. If there is interest in scattering between widely separated (in ${\mathbf k}$) multiple conduction minima then the applicability of these calculations will depend on the importance of possible second-order corrections to the expansion in Eq.~(\ref{Expansion}). Consideration of these effects is beyond the scope of this publication, although we note that the expansion in Eq.~(\ref{Expansion}) could in principle be extended to higher-order polynomials in ${\mathbf k}$ to describe these effects. 

The scattering amplitude between two  states in a single conduction band, in the Born approximation, is the matrix element of  the effective potential between simple plane wave states -- the periodic parts of the wave functions having  already been incorporated in $\tilde V$.
For the case of a $\kv$-independent  potential:  
\be
\tilde V_{\kv'\alpha',\kv\alpha}=V_0\delta_{\alpha'\alpha}+iV_0\sum_{ijk} \lambda_{ijk} k_i k'_j [\sigma_k ]_{\alpha'\alpha}\ .
\ee
 The transition rate  between states $\kv \alpha$ and $\kv'\alpha'$ is then
\al{ \label{eq:goldenrule}
  P(\kv\alpha ;\kv'\alpha')=\frac{2\pi}{\hbar}|\tilde V_{\kv'\alpha',\kv\alpha}|^2\delta(E_{c\v k'}-E_{c\v k}).
}

The scattering potential $V_0$ mimics the scattering that produces the  experimental carrier mobility $\mu={e \tau_p}/{m^*}$, where $m^*$ is the effective mass of the band near  $\Gamma$,     
\al{\label{scatt}
{\tau_p^{-1}}=
{\sum_{\kv\alpha,\kv'\alpha'}P(\kv\alpha; \kv'\alpha')f_\kv(1-f_{\kv'})}/{\sum_{\kv,\kv'}f_\kv(1-f_{\kv'})}
}
is the momentum relaxation rate and $f_\kv$ is the  Fermi-Dirac equilibrium distribution function.  The expression for the spin lifetime $\tau_s$ is like Eq.~(\ref{scatt}), but with the sum over $\alpha'$ restricted to $\alpha'= -\alpha$,
\al{\label{spinscatt}
{\tau_s^{-1}}=
{\sum_{\kv\alpha,\kv'}P(\kv\alpha; \kv'-\alpha)f_\kv(1-f_{\kv'})}/{\sum_{\kv,\kv'}f_\kv(1-f_{\kv'})}.
}Spin flips occur via  mixing of different spin states into the wave functions of eigenstates of different momenta, which produces spin flips as the carriers scatter from interactions with impurities and phonons. 

Equations~(\ref{eq:effpot})-(\ref{spinscatt}) are principal results of the formalism presented here. Although formally $\tau_s$ is a pseudospin lifetime, if the doubly-degenerate states at $\kv_\mu$ can be written as unentangled product states of orbit and spin then $\tau_s$ can be identified as the actual spin lifetime. This occurs for the $s$-orbital conduction band of III-V semiconductors and the $d_{xy}$-orbital conduction band of strained STO or LAO/STO. We now verify results obtained from these equations for III-V semiconductors, and then apply the results to STO-based materials.

\section{III-V semiconductors} 

A simple \kp model of the electronic structure near zone center ${\bf k_\mu}=0$ incorporating eight bands and spin-orbit interaction can be analytically evaluated for $\lambda_{ijk}$. The Hamiltonian is
\begin{equation}
H_{\mathbf k} = H_{\mathbf k=0} + \frac{\hbar {\mathbf k}\cdot {\mathbf P}}{m}
\end{equation}
where $m$ is the electron's free mass and ${\mathbf P}$ is the momentum operator. The free kinetic energy of the electron is neglected. In the eight-band model the eigenstates of $H_{\mathbf k=0}$ correspond to the conduction band spin up and down states as well as heavy, light and split-off holes with spin up and down. The Hamiltonian  for this set of basis states 
\begin{widetext}
\begin{equation}
H_{\mathbf k} =\left[\begin{array}{cccccccc}
E_g&0&\frac{i\hbar P}{\sqrt{2}m}{k_+}&0
&\frac{i\sqrt{2}\hbar P}{\sqrt{3}m}k_z &\frac{i\hbar P}{\sqrt{6}m}k_-
&\frac{i\hbar P}{\sqrt{3}m}k_z &\frac{i\hbar P}{\sqrt{3}m}k_-\\
0&E_g&0&\frac{i\hbar P}{\sqrt{2}m}{k_-}
&\frac{i\hbar P}{\sqrt{6}m}k_+&\frac{i\sqrt{2}\hbar P}{\sqrt{3}m}k_z
&\frac{i\hbar P}{\sqrt{3}m}k_+&\frac{i\hbar P}{\sqrt{3}m}k_z\\
-\frac{i\hbar P}{\sqrt{2}m}{k_-}&0&0&0&0&0&0&0\\
0&-\frac{i\hbar P}{\sqrt{2}m}{k_+}&0&0&0&0&0&0\\
-\frac{i\sqrt{2}\hbar P}{\sqrt{3}m}k_z&-\frac{i\hbar P}{\sqrt{6}m}k_-&0&0&0&0&0&0\\
-\frac{i\hbar P}{\sqrt{6}m}k_+&-\frac{i\sqrt{2}\hbar P}{\sqrt{3}m}k_z&0&0&0&0&0&0\\
-\frac{i\hbar P}{\sqrt{3}m}k_z &-\frac{i\hbar P}{\sqrt{3}m}k_-&0&0&0&0&-\Delta&0\\
-\frac{i\hbar P}{\sqrt{3}m}k_+&-\frac{i\hbar P}{\sqrt{3}m}k_z&0&0&0&0&0&-\Delta\\

\end{array}
\right]
\end{equation}
\end{widetext}
where $k_+ = k_x+ik_y$ and $k_- = k_x - ik_y$, E$_g$ is the band gap, $\Delta$ the spin orbit splitting in the the valence bands, and $P$ the magnitude of the momentum matrix element between conduction and valence bands\cite{Cardona1988}. Evaluation of Eq.~(\ref{eq:so-tensor}) for this Hamiltonian  results in $\lambda_{ijk}=\lambda\epsilon_{ijk}$ and the analytic expression 
\begin{equation}
\lambda=\frac{\hbar^2P^2}{3m^{*2}}\left[\frac{1}{E_g^{2}}-\frac{1}{(E_g+\Delta)^{2}}\right].\label{lambdakp}
\end{equation}

Eq.~(\ref{eq:so-tensor}) can also be straightforwardly evaluated for any tight-binding Hamiltonian, which are expressed as Hamiltonians between Bloch sums, labeled by orbital and atomic site\cite{Yu3ed}. The ${\mathbf k}$-dependent terms that appear in such Hamiltonians originate from overlap matrix elements between neighbors, and generally have the form 
\begin{equation}
\sum_{\mathbf d_n} {\rm e}^{i {\mathbf k}\cdot{\mathbf d_n}}\label{bsform}
\end{equation}
where the ${\mathbf d_n}$ run over the distances between neighboring atoms coupled by the overlap matrix elements. Derivatives of terms such as those appearing in Eq.~(\ref{bsform}) with respect to ${\mathbf k}$ are simple to evaluate.  The \kp expression from Eq.~(\ref{lambdakp}) and the $\lambda$ computed from an  spds$^*$ tight-binding Hamiltonian obtained from Ref.~\onlinecite{Jancu1998} agree, as shown in Table~\ref{tab:SOIcomparison}:

\begin{table}[h]
\centering
\begin{tabular}{|l|cccr|}
\hline
\textrm{Method}&
\textrm{GaAs}&
\textrm{InP}&
\textrm{GaSb}&
\textrm{InSb}\\
\hline
$\v k \cdot \v p$ & 4.4 & 1.7  & 32.5 & 544.1\\
Tight-binding & 4.6 & 1.8 & 34.6 & 583.8\\
From Eq.~(\ref{eq:lambdacomparison}) & 5.1 & 1.7 & 39.7 &630.9 \\
\hline
\end{tabular}
\caption{\label{tab:SOIcomparison}
Spin-orbit interaction parameter $\lambda$ in  units of \AA$^2$ }
\end{table}

A check of our spin lifetime is provided by an analytical expression derived from an eight-band \kp model for the ratio of the spin lifetimes calculated for a $\kv$-independent potential from the Elliott-Yafet mechanism and  the momentum relaxation time\cite{Meier1984},
\al{\label{eq:opticalorientation}
\frac{\tau_p}{\tau_s}=\frac{32}{81} 
\left(\frac{1}{E_g}\right )^2\eta^2 \left (\frac{1-\eta /2}{1-\eta /3}\right )^2 E_k^2,
}
where $E_g$ is the band gap,  $\eta=\Delta /(E_g+\Delta)$, and $E_k=\hbar^2 k^2 /2m^{*}$.
The ratio from Eq.~(\ref{scatt}) is:
\al{
\frac{\tau_p}{\tau_s}=\lambda^2\frac{8{m^*}^2}{3\hbar^4} E_k^2
}
which has the same functional form. If
\al{\label{eq:lambdacomparison}
\lambda=\frac{\hbar^2}{2m^*}\frac{4}{3\sqrt 3}\frac{1}{E_g} \eta \left (\frac{1-\eta /2}{1-\eta /3}\right ),
}
then the two expressions agree. We report in Table \ref{tab:SOIcomparison} the implied value of $\lambda$ from Eq.~(\ref{eq:lambdacomparison}), indicating good agreement between our formalism and previously-obtained results for spin lifetimes in III-V semiconductors. Experimental spin lifetimes in such materials are not useful for direct comparison, as they are dominated by effects absent in STO and other inversion-symmetric materials\cite{Meier1984}.

\section{Strontium Titanate based materials}   

\begin{figure}
\includegraphics[width=3.2in]{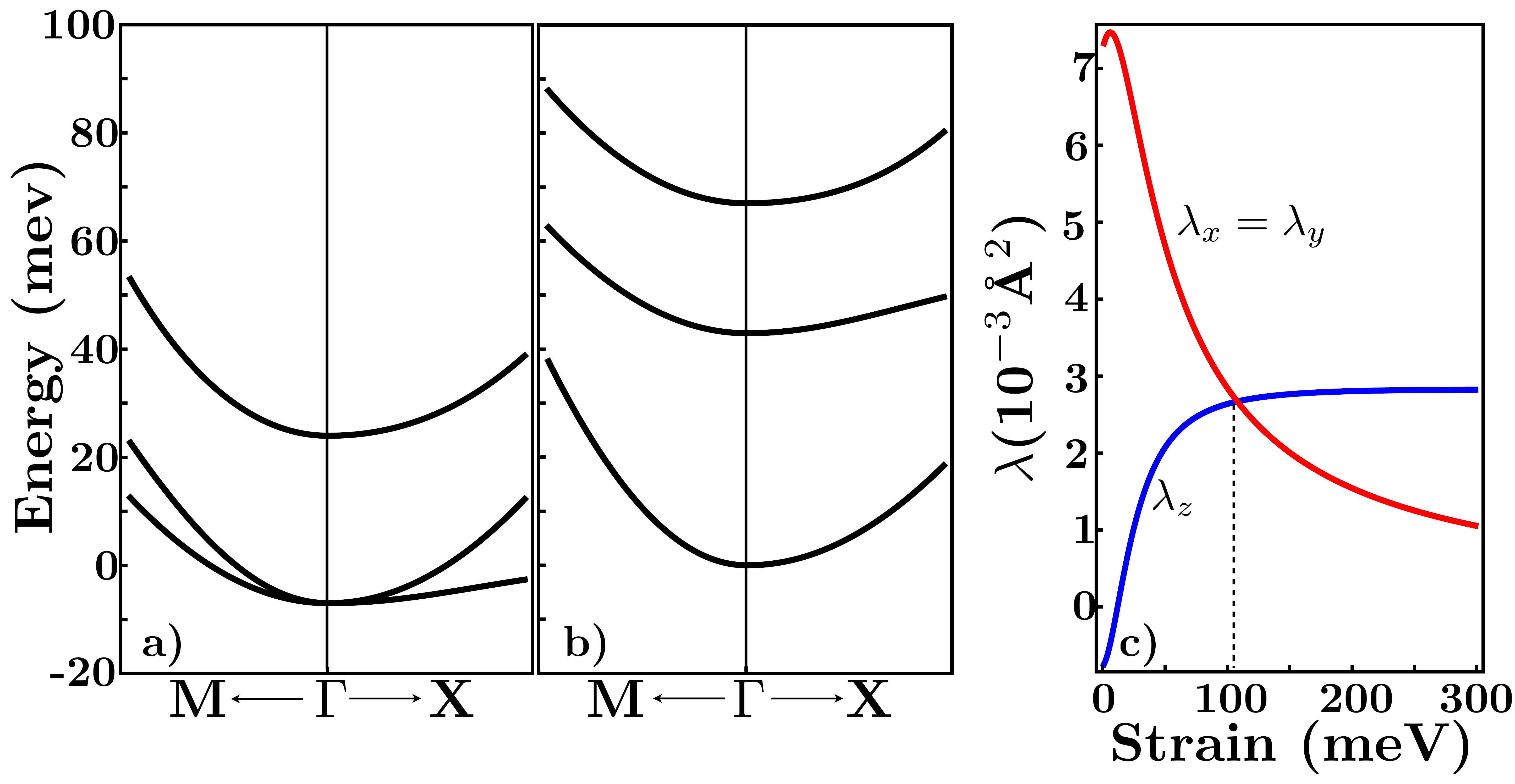}
\caption{Conduction bands of STO calculated by the tight-binding method described in the text. (10\% of the Brillouin zone in each direction is shown) (a) \textit{unstrained:} the lowest conduction band at  $\Gamma$  is four-fold degenerate (b) \textit{strained:} a compressive uniaxial stress induces a splitting (here 50~meV) that splits the degeneracy and results in three doubly degenerate conduction bands. (c) Magnitude of spin-orbit interaction $\lambda_k$ as a function of the conduction band splitting at $\Gamma$  due to strain or confinement.  This formulation is not applicable to the case of zero strain due to the four-fold degeneracy at the $\Gamma$ point.}
\label{fig:bands-lambda}
\end{figure}

For STO there exists only one momentum corresponding to the conduction band minimum, and the electronic states near this minimum at the Brillouin zone center mostly consist of Ti \textit{d}-orbitals. The crystal potential splits these conduction bands into sixfold t$_{2g}$ bands (d$_{xy}$, d$_{yz}$, d$_{zx}$) and fourfold (higher-energy) e$_g$ bands (d$_{x^2-y^2}$, d$_{3z^2 -r^2}$); spin-orbit coupling results in a further splitting ($\approx$ 30 meV) of the lower t$_{2g}$ bands into fourfold and and twofold bands, as shown in Fig.~\ref{fig:bands-lambda}(a).  We consider strained STO, in which the compressive strain breaks the fourfold degeneracy at the $\Gamma$-point and results in well-resolved, doubly degenerate subbands in the plane perpendicular to the growth direction, as shown in Fig.~\ref{fig:bands-lambda}(b) for a splitting of $\sim 50$~meV. The same energy splitting is produced by an interface and leads to the electronic structure of the LAO/STO 2DEG\cite{Salluzzo2009}.  

The electronic structure is calculated using a tight-binding Hamiltonian with values  from Ref.~\onlinecite{Kahn1964}; the parametrization omits \textit{s}-orbitals of strontium and includes nearest-neighbor interactions between \textit{2p}-orbitals of oxygen and full \textit{3d}-orbitals of titanium as opposed to simpler parameterizations with only t$_{2g}$ bands, such as in Ref.~\onlinecite{Mattheiss1972a} and Ref.~\onlinecite{Wolfram1972}.  The spin-orbit couplings, absent in Ref.~\onlinecite{Kahn1964}, are computed from atomic spectra tables\cite{Moore12}.
This results in a 30 meV spin-orbit splitting, in agreement with first principle calculations\cite{Marel2011}.
Here the Rashba spin splittings induced by the effective confinement fields along the growth direction at the interface are ignored; these splittings further reduce the spin lifetimes, 
thus our results can be viewed as the long spin lifetimes obtainable if the confinement field that induces the Rashba spin splitting has been compensated by another field, such as a gate field\cite{Lau2005}.

There are only six non-zero elements of $\lambda_{ijk}$ from Eq.~(\ref{eq:so-tensor}) at the minimum of the conduction band (\m{\Gamma} point) for STO $\lambda_{ijk}=-\lambda_{jik}=\epsilon_{ij}\lambda_k$, where $i$, $j$, and $k$  all differ. From our tight-binding band structure of SrTiO$_3$, and taking $z$ the direction of a uniaxial strain,  $\lambda_x=\lambda_y= 0.0047$~\AA $^2$ and $\lambda_z= 0.0021$~\AA$^2$  for a strain resulting in 50 meV splitting in the conduction band minimum. The dependence of $\lambda$ on the strain is shown in Fig.~\ref{fig:bands-lambda}(c). Large strain destroys $\lambda_x$ and $\lambda_y$ and leaves $\lambda_z$ constant at 0.0028 \AA$^2$. The strain value where $\lambda_x=\lambda_y=\lambda_z$ is around 110 meV, and the lowest conduction band ($d_{xy}$-like) has isotropic dispersion in the $xy$ plane.\footnote{Below a temperature of 100K STO undergoes a second-order phase transition from cubic to tetragonal structure while oxygens in STO start to rotate. \cite{Mattheiss1972a}. This rotation breaks the cubic symmetry and causes a further shift in the higher conduction bands, which we neglect here.} These values of $\lambda$ are approximately three orders of magnitude smaller than those for III-V semiconductors, which will lead to correspondingly longer spin coherence times (proportional to $\lambda^{-2}$).

\begin{figure}
\includegraphics[width=3.2in]{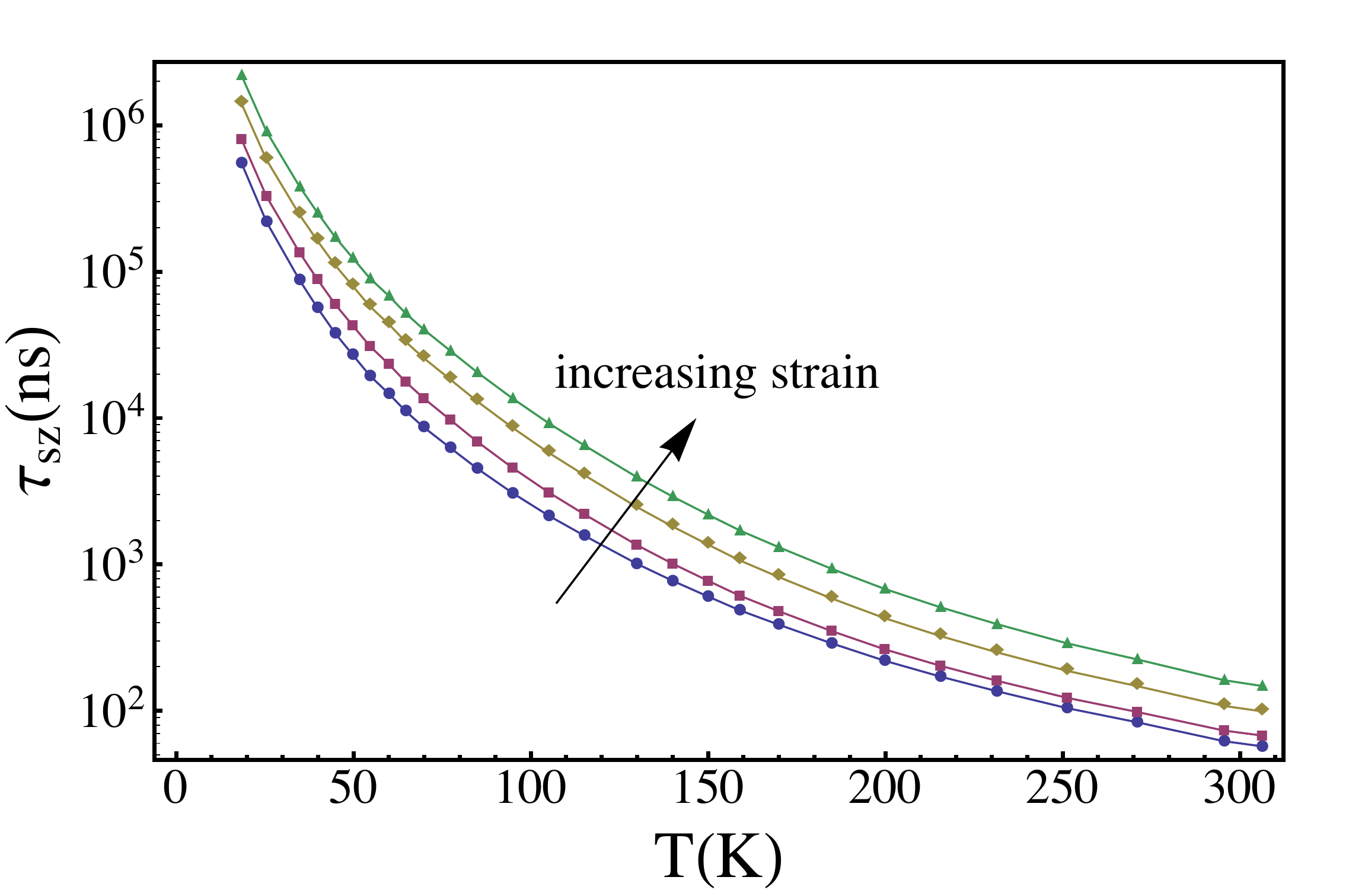}
\caption{Spin relaxation time of bulk strontium titanate  as a function of temperature and strain. The carrier concentration is $1.0\times 10^{18}$cm$^{-3}$ and the mobility varies from 5-7000 cm$^2$V$^{-1}$s$^{-1}$ as reported in Ref.~\onlinecite{Moos1995}. }
\label{fig:sto}
\end{figure}

Spin lifetimes for bulk strained strontium titanate for spin parallel to $\hat z$ ($\tau_{sz}$, Fig.~\ref{fig:sto}) were evaluated from Eqs.~(\ref{eq:goldenrule})-(\ref{scatt}) using reported\cite{Moos1995} carrier mobilities and densities. 
Spins oriented along $\hat x$ or $\hat y$ exhibit the same lifetime dependence on  temperature and strain, but  are shorter by $\sim 15\%$ at low temperatures and $\sim 10\%$ at room temperature from $\tau_{sz}$. Strain splitting of the bands is increased uniformly from 50 meV to 110 meV which reduces the spin mixing of these bands, resulting in a  longer spin lifetime.

\begin{figure}
\includegraphics[width=3.2in]{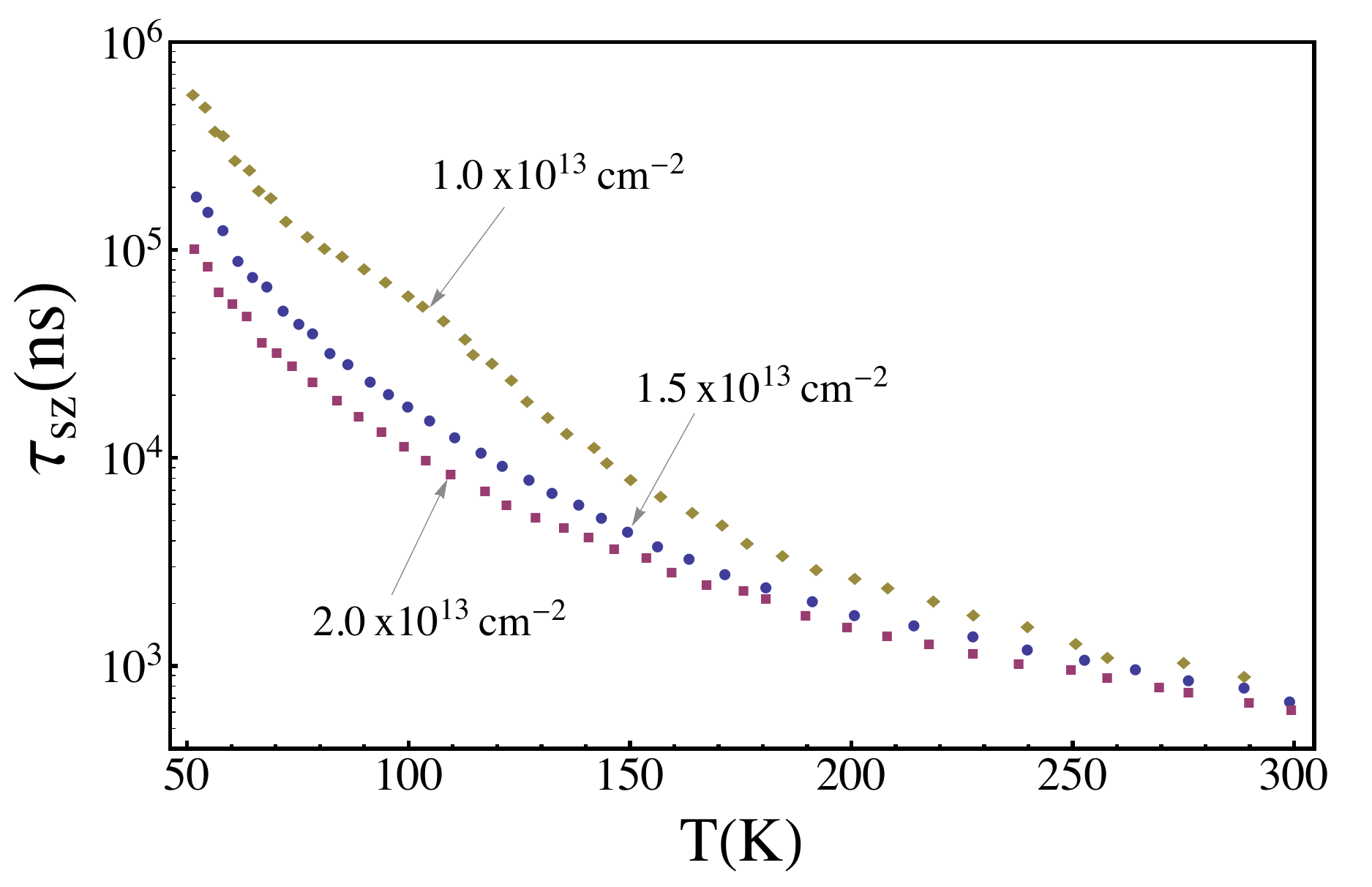}
\caption{(a) Spin relaxation time as a function of temperature for three densities of carriers in the LAO/STO 2DEG.  
The mobilities and densities correspond to those reported in Ref.~\onlinecite{Kalabukhov2007}.}
\label{fig:lao-sto}
\end{figure}

Our calculated spin relaxation times for a LAO/STO 2DEG are shown in Fig.~\ref{fig:lao-sto}, for several experimentally achieved carrier densities (corresponding to several oxygen partial pressures during growth). The dominant source of the reduction of carrier spin lifetime with temperature is an increase in the scattering rate from phonons at higher temperatures. These spin lifetimes greatly exceed those of bulk III-V semiconductors at room temperature, and are one to two orders of magnitude longer than room-temperature spin lifetimes in specially-designed GaAs quantum wells grown along the [110] direction\cite{Karimov2003}.  The resulting spin lifetimes are of the same order as those of the strained STO at low temperatures, but one order of magnitude greater at room temperature.

\section{Conclusions}

This systematic approach to the calculation of the effective spin-orbit interaction and the Elliot-Yafet spin relaxation rate in doubly-degenerate bands is broadly applicable to centro-symmetric nonmagnetic materials. Starting from a calculated band structure we have derived a compact, gauge invariant formula for the spin-orbit interaction tensor, and applied it to spin lifetimes. These results reproduce previous calculations via $\kv\cdot{\bf p}$ theory of spin lifetimes in III-V semiconductors. Our results also support the presence of robust, room-temperature spin dynamics in oxide materials such as STO and the LAO/STO interfacial 2DEG. As centro-symmetric materials have recently taken up a more prominent role in spin-dependent phenomena (e.g. large spin Hall effects in cubic metals, spin lifetimes in diamond-based materials) it is expected that this approach will apply to a broad range of materials and spin-dependent phenomena.  

\begin{acknowledgments}
We acknowledge support by an ARO MURI.

\end{acknowledgments}
\appendix
\section{Structure of the intra-band connection matrix}\label{appA}
In this section we prove that the intra-band connection matrix
\be
\mathbf A_{n \alpha, n \beta}\equiv i \braket{u_{n \beta}}{\partial{u_{n \tilde{\mathbf k} \alpha}}/\partial{\tilde {\mathbf k}}}_{\tilde{\mathbf k} =0}\,,
\ee
where $\alpha$ and $\beta$ are pseudospin indices with values $\pm 1$, has the following properties
\be\label{e1}
\mathbf A_{n \alpha, n \alpha}=-\mathbf A_{n -\alpha, n -\alpha}
\ee
and
\be\label{e2}
\mathbf A_{n \alpha, n -\alpha}= 0\,.
\ee
To see this we explicitly write down the matrix element
\begin{widetext}
\be
\mathbf A_{n \alpha, n \beta} = i\sum_{\tau,\tau',\tau''}\int d\rv \left([i\sigma_y]_{\tau\tau'}u^*_{n\tilde \kv-\beta}(-\rv,\tau')\right)^* \frac{\partial}{\partial \tilde \kv} \left([i\sigma_y]_{\tau\tau''}u^*_{n\tilde \kv-\alpha}(-\rv,\tau'')   \right)\,,
\ee
\end{widetext}
where  we have explicitly denoted the pseudospin components of the spinor $u_{n\tilde \kv}(\rv)$ as $u_{n\tilde \kv}(r,\tau)$, with $\tau=\pm 1$, and we have used Eq. (1) of the main text, to express $u_{n\tilde \kv\alpha}$ in terms of $u^*_{n\tilde \kv -\alpha}$.  Carrying out first the sum over $\tau$ with $\sum_\tau [\sigma_y^*]_{\tau\tau'}[\sigma_y]_{\tau\tau''} =
\sum_\tau [\sigma_y]_{\tau'\tau}[\sigma_y]_{\tau\tau''} = \delta_{\tau'\tau''}$ we obtain
\be
\mathbf A_{n \alpha, n \beta} = i\sum_{\tau'}\int d\rv u_{n\tilde \kv-\beta}(-\rv,\tau')\frac{\partial}{\partial \tilde \kv} u^*_{n\tilde \kv-\alpha}(-\rv,\tau') \,.
\ee
We change integration variable from $\rv$ to $-\rv$ and transfer the operator $\frac{\partial}{\partial \tilde \kv}$ from the right to the left wave function with a change of sign (this is allowed because   $\frac{\partial}{\partial \tilde \kv} \langle u_ {n\tilde \kv-\alpha}|u_{n\tilde \kv-\beta}\rangle =\frac{\partial}{\partial \tilde \kv}\delta_{\alpha\beta}=0$).  So we arrive at
\al{
\mathbf A_{n \alpha, n \beta} &= -i\sum_{\tau'}\int d\rv u^*_{n\tilde \kv-\alpha}(\rv,\tau')\frac{\partial}{\partial \tilde \kv} u_{n\tilde \kv-\beta}(\rv,\tau')\\
&=- \mathbf A_{n -\beta, n -\alpha}\,.
}
Setting $\beta=\alpha$ in the above equation yields Eq.~(\ref{e1}).  Setting $\beta=-\alpha$  yields $\mathbf A_{n \alpha, n -\alpha}=- \mathbf A_{n \alpha, n -\alpha}$, which implies Eq.~(\ref{e2}). 

\section{Derivation of Eq.~(\ref{matrixelement})} \label{appB}
We begin by rewriting Eq.~(\ref{matrixelement}) explicitly as follows
\be
V_{c\tilde \kv'\alpha',c\tilde \kv\alpha}=\sum_\sigma\int d\rv e^{i(\tilde \kv-\tilde \kv')\cdot\rv}u^*_{c\tilde \kv'\alpha'}(\rv,\sigma) V(\rv)u_{c\tilde \kv\alpha}(\rv,\sigma)\,,
\ee
where the normalization volume has been set to $1$.
The periodic wave functions $u_{c\tilde \kv\alpha}(\rv,\sigma)$ are expanded to first order in $\tilde \kv$ according to Eq.~\ref{Expansion}.  The integral over space is rewritten as a sum of integrals over unit cells, $\Omega({\bf R})$,  centered at lattice sites ${\bf R}$:
\be
\int d\rv = \sum_{\bf R} \int_{\Omega({\bf R})} d\rv\,.
\ee 
 Within each unit cell the potential and the exponential factor are regarded as constants  equal to $V({\bf R})$ and $e^{i(\tilde \kv-\tilde \kv')\cdot{\bf R}}$ respectively.   The remaining integration over the periodic part of the Bloch wave functions is done with the help of the orthonormality relations
\be
 \sum_{\sigma}\int_{\Omega({\bf R})} d\rv u^*_{n'\alpha'}(\rv,\sigma) u_{n\alpha}(\rv,\sigma)  =\frac{1}{N}\delta_{nn'}\delta_{\alpha\alpha'}\,,
 \ee
 where $N$ is the number of unit cells, each unit cell having a volume $\frac{1}{N}$ in units in which the total volume is $1$.  Lastly, the sum over ${\bf R}$  of a slowly varying function $f({\bf R})$ is replaced by an integral over the whole space:
 \be
 \frac{1}{N} \sum_{\bf R} f({\bf R}) =  \int d\rv f(\rv)\,.
 \ee
 By following this procedure Eq.~\ref{matrixelement} is easily obtained.

\bibliography{central-bibliography}

\begin{thebibliography}{33}%
\makeatletter
\providecommand \@ifxundefined [1]{%
 \@ifx{#1\undefined}
}%
\providecommand \@ifnum [1]{%
 \ifnum #1\expandafter \@firstoftwo
 \else \expandafter \@secondoftwo
 \fi
}%
\providecommand \@ifx [1]{%
 \ifx #1\expandafter \@firstoftwo
 \else \expandafter \@secondoftwo
 \fi
}%
\providecommand \natexlab [1]{#1}%
\providecommand \enquote  [1]{``#1''}%
\providecommand \bibnamefont  [1]{#1}%
\providecommand \bibfnamefont [1]{#1}%
\providecommand \citenamefont [1]{#1}%
\providecommand \href@noop [0]{\@secondoftwo}%
\providecommand \href [0]{\begingroup \@sanitize@url \@href}%
\providecommand \@href[1]{\@@startlink{#1}\@@href}%
\providecommand \@@href[1]{\endgroup#1\@@endlink}%
\providecommand \@sanitize@url [0]{\catcode `\\12\catcode `\$12\catcode
  `\&12\catcode `\#12\catcode `\^12\catcode `\_12\catcode `\%12\relax}%
\providecommand \@@startlink[1]{}%
\providecommand \@@endlink[0]{}%
\providecommand \url  [0]{\begingroup\@sanitize@url \@url }%
\providecommand \@url [1]{\endgroup\@href {#1}{\urlprefix }}%
\providecommand \urlprefix  [0]{URL }%
\providecommand \Eprint [0]{\href }%
\providecommand \doibase [0]{http://dx.doi.org/}%
\providecommand \selectlanguage [0]{\@gobble}%
\providecommand \bibinfo  [0]{\@secondoftwo}%
\providecommand \bibfield  [0]{\@secondoftwo}%
\providecommand \translation [1]{[#1]}%
\providecommand \BibitemOpen [0]{}%
\providecommand \bibitemStop [0]{}%
\providecommand \bibitemNoStop [0]{.\EOS\space}%
\providecommand \EOS [0]{\spacefactor3000\relax}%
\providecommand \BibitemShut  [1]{\csname bibitem#1\endcsname}%
\let\auto@bib@innerbib\@empty
\bibitem [{\citenamefont {Balasubramanian}\ \emph {et~al.}(2009)\citenamefont
  {Balasubramanian}, \citenamefont {Neumann}, \citenamefont {Twitchen},
  \citenamefont {Markham}, \citenamefont {Kolesov}, \citenamefont {Mizuochi},
  \citenamefont {Isoya}, \citenamefont {Achard}, \citenamefont {Beck},
  \citenamefont {Tissler}, \citenamefont {Jacques}, \citenamefont {Hemmer},
  \citenamefont {Jelezko},\ and\ \citenamefont
  {Wrachtrup}}]{Balasubramanian2009}%
  \BibitemOpen
  \bibfield  {author} {\bibinfo {author} {\bibfnamefont {G.}~\bibnamefont
  {Balasubramanian}}, \bibinfo {author} {\bibfnamefont {P.}~\bibnamefont
  {Neumann}}, \bibinfo {author} {\bibfnamefont {D.}~\bibnamefont {Twitchen}},
  \bibinfo {author} {\bibfnamefont {M.}~\bibnamefont {Markham}}, \bibinfo
  {author} {\bibfnamefont {R.}~\bibnamefont {Kolesov}}, \bibinfo {author}
  {\bibfnamefont {N.}~\bibnamefont {Mizuochi}}, \bibinfo {author}
  {\bibfnamefont {J.}~\bibnamefont {Isoya}}, \bibinfo {author} {\bibfnamefont
  {J.}~\bibnamefont {Achard}}, \bibinfo {author} {\bibfnamefont
  {J.}~\bibnamefont {Beck}}, \bibinfo {author} {\bibfnamefont {J.}~\bibnamefont
  {Tissler}}, \bibinfo {author} {\bibfnamefont {V.}~\bibnamefont {Jacques}},
  \bibinfo {author} {\bibfnamefont {P.~R.}\ \bibnamefont {Hemmer}}, \bibinfo
  {author} {\bibfnamefont {F.}~\bibnamefont {Jelezko}}, \ and\ \bibinfo
  {author} {\bibfnamefont {J.}~\bibnamefont {Wrachtrup}},\ }\href {\doibase
  10.1038/NMAT2420} {\bibfield  {journal} {\bibinfo  {journal} {Nature
  Materials}\ }\textbf {\bibinfo {volume} {8}},\ \bibinfo {pages} {383}
  (\bibinfo {year} {2009})}\BibitemShut {NoStop}%
\bibitem [{\citenamefont {Koehl}\ \emph {et~al.}(2011)\citenamefont {Koehl},
  \citenamefont {Buckley}, \citenamefont {Heremans}, \citenamefont {Calusine},\
  and\ \citenamefont {Awschalom}}]{Koehl2011}%
  \BibitemOpen
  \bibfield  {author} {\bibinfo {author} {\bibfnamefont {W.~F.}\ \bibnamefont
  {Koehl}}, \bibinfo {author} {\bibfnamefont {B.~B.}\ \bibnamefont {Buckley}},
  \bibinfo {author} {\bibfnamefont {F.~J.}\ \bibnamefont {Heremans}}, \bibinfo
  {author} {\bibfnamefont {G.}~\bibnamefont {Calusine}}, \ and\ \bibinfo
  {author} {\bibfnamefont {D.~D.}\ \bibnamefont {Awschalom}},\ }\href@noop {}
  {\bibfield  {journal} {\bibinfo  {journal} {Nature}\ }\textbf {\bibinfo
  {volume} {479}},\ \bibinfo {pages} {84} (\bibinfo {year} {2011})}\BibitemShut
  {NoStop}%
\bibitem [{\citenamefont {Han}\ \emph {et~al.}(2013)\citenamefont {Han},
  \citenamefont {Jiang}, \citenamefont {Kajdos}, \citenamefont {Yang},
  \citenamefont {Stemmer},\ and\ \citenamefont {Parkin}}]{Han2013}%
  \BibitemOpen
  \bibfield  {author} {\bibinfo {author} {\bibfnamefont {W.}~\bibnamefont
  {Han}}, \bibinfo {author} {\bibfnamefont {X.}~\bibnamefont {Jiang}}, \bibinfo
  {author} {\bibfnamefont {A.}~\bibnamefont {Kajdos}}, \bibinfo {author}
  {\bibfnamefont {S.-H.}\ \bibnamefont {Yang}}, \bibinfo {author}
  {\bibfnamefont {S.}~\bibnamefont {Stemmer}}, \ and\ \bibinfo {author}
  {\bibfnamefont {S.~S.~P.}\ \bibnamefont {Parkin}},\ }\href@noop {} {\bibfield
   {journal} {\bibinfo  {journal} {Nature Comm.}\ }\textbf {\bibinfo {volume}
  {4}},\ \bibinfo {pages} {2134} (\bibinfo {year} {2013})}\BibitemShut
  {NoStop}%
\bibitem [{\citenamefont {Caviglia}\ \emph {et~al.}(2010)\citenamefont
  {Caviglia}, \citenamefont {Gabay}, \citenamefont {Gariglio}, \citenamefont
  {Reyren}, \citenamefont {Cancellieri},\ and\ \citenamefont
  {Triscone}}]{Caviglia2010}%
  \BibitemOpen
  \bibfield  {author} {\bibinfo {author} {\bibfnamefont {A.~D.}\ \bibnamefont
  {Caviglia}}, \bibinfo {author} {\bibfnamefont {M.}~\bibnamefont {Gabay}},
  \bibinfo {author} {\bibfnamefont {S.}~\bibnamefont {Gariglio}}, \bibinfo
  {author} {\bibfnamefont {N.}~\bibnamefont {Reyren}}, \bibinfo {author}
  {\bibfnamefont {C.}~\bibnamefont {Cancellieri}}, \ and\ \bibinfo {author}
  {\bibfnamefont {J.-M.}\ \bibnamefont {Triscone}},\ }\href {\doibase
  10.1103/PhysRevLett.104.126803} {\bibfield  {journal} {\bibinfo  {journal}
  {Phys. Rev. Lett.}\ }\textbf {\bibinfo {volume} {104}},\ \bibinfo {pages}
  {126803} (\bibinfo {year} {2010})}\BibitemShut {NoStop}%
\bibitem [{\citenamefont {Reyren}\ \emph {et~al.}(2012)\citenamefont {Reyren},
  \citenamefont {Bibes}, \citenamefont {Lesne}, \citenamefont {George},
  \citenamefont {Deranlot}, \citenamefont {Collin}, \citenamefont
  {Barth\'el\'emy},\ and\ \citenamefont {Jaffr\`es}}]{Reyren2012}%
  \BibitemOpen
  \bibfield  {author} {\bibinfo {author} {\bibfnamefont {N.}~\bibnamefont
  {Reyren}}, \bibinfo {author} {\bibfnamefont {M.}~\bibnamefont {Bibes}},
  \bibinfo {author} {\bibfnamefont {E.}~\bibnamefont {Lesne}}, \bibinfo
  {author} {\bibfnamefont {J.-M.}\ \bibnamefont {George}}, \bibinfo {author}
  {\bibfnamefont {C.}~\bibnamefont {Deranlot}}, \bibinfo {author}
  {\bibfnamefont {S.}~\bibnamefont {Collin}}, \bibinfo {author} {\bibfnamefont
  {A.}~\bibnamefont {Barth\'el\'emy}}, \ and\ \bibinfo {author} {\bibfnamefont
  {H.}~\bibnamefont {Jaffr\`es}},\ }\href {\doibase
  10.1103/PhysRevLett.108.186802} {\bibfield  {journal} {\bibinfo  {journal}
  {Phys. Rev. Lett.}\ }\textbf {\bibinfo {volume} {108}},\ \bibinfo {pages}
  {186802} (\bibinfo {year} {2012})}\BibitemShut {NoStop}%
\bibitem [{\citenamefont {Jalan}\ \emph {et~al.}(2011)\citenamefont {Jalan},
  \citenamefont {Allen}, \citenamefont {Beltz}, \citenamefont {Moetakef},\ and\
  \citenamefont {Stemmer}}]{Jalan2011}%
  \BibitemOpen
  \bibfield  {author} {\bibinfo {author} {\bibfnamefont {B.}~\bibnamefont
  {Jalan}}, \bibinfo {author} {\bibfnamefont {S.~J.}\ \bibnamefont {Allen}},
  \bibinfo {author} {\bibfnamefont {G.~E.}\ \bibnamefont {Beltz}}, \bibinfo
  {author} {\bibfnamefont {P.}~\bibnamefont {Moetakef}}, \ and\ \bibinfo
  {author} {\bibfnamefont {S.}~\bibnamefont {Stemmer}},\ }\href {\doibase
  10.1063/1.3571447} {\bibfield  {journal} {\bibinfo  {journal} {Applied
  Physics Letters}\ }\textbf {\bibinfo {volume} {98}},\ \bibinfo {pages}
  {132102} (\bibinfo {year} {2011})}\BibitemShut {NoStop}%
\bibitem [{\citenamefont {Son}\ \emph {et~al.}(2010)\citenamefont {Son},
  \citenamefont {Moetakef}, \citenamefont {Jalan}, \citenamefont {Bierwagen},
  \citenamefont {Wright}, \citenamefont {Engel-Herbert},\ and\ \citenamefont
  {Stemmer}}]{Son2010}%
  \BibitemOpen
  \bibfield  {author} {\bibinfo {author} {\bibfnamefont {J.}~\bibnamefont
  {Son}}, \bibinfo {author} {\bibfnamefont {P.}~\bibnamefont {Moetakef}},
  \bibinfo {author} {\bibfnamefont {B.}~\bibnamefont {Jalan}}, \bibinfo
  {author} {\bibfnamefont {O.}~\bibnamefont {Bierwagen}}, \bibinfo {author}
  {\bibfnamefont {N.~J.}\ \bibnamefont {Wright}}, \bibinfo {author}
  {\bibfnamefont {R.}~\bibnamefont {Engel-Herbert}}, \ and\ \bibinfo {author}
  {\bibfnamefont {S.}~\bibnamefont {Stemmer}},\ }\href {\doibase
  10.1038/nmat2750} {\bibfield  {journal} {\bibinfo  {journal} {Nature
  materials}\ }\textbf {\bibinfo {volume} {9}},\ \bibinfo {pages} {482}
  (\bibinfo {year} {2010})}\BibitemShut {NoStop}%
\bibitem [{\citenamefont {Ohtomo}\ and\ \citenamefont
  {Hwang}(2004)}]{Ohtomo2004}%
  \BibitemOpen
  \bibfield  {author} {\bibinfo {author} {\bibfnamefont {A.}~\bibnamefont
  {Ohtomo}}\ and\ \bibinfo {author} {\bibfnamefont {H.~Y.}\ \bibnamefont
  {Hwang}},\ }\href {\doibase 10.1038/nature02308} {\bibfield  {journal}
  {\bibinfo  {journal} {Nature}\ }\textbf {\bibinfo {volume} {427}},\ \bibinfo
  {pages} {423} (\bibinfo {year} {2004})}\BibitemShut {NoStop}%
\bibitem [{\citenamefont {Meier}\ and\ \citenamefont
  {Zachachrenya}(1984)}]{Meier1984}%
  \BibitemOpen
  \bibfield  {author} {\bibinfo {author} {\bibfnamefont {F.}~\bibnamefont
  {Meier}}\ and\ \bibinfo {author} {\bibfnamefont {B.~P.}\ \bibnamefont
  {Zachachrenya}},\ }\href@noop {} {\emph {\bibinfo {title} {Optical
  Orientation: Modern Problems in Condensed Matter Science}}},\ Vol.~\bibinfo
  {volume} {8}\ (\bibinfo  {publisher} {North-Holland},\ \bibinfo {address}
  {Amsterdam},\ \bibinfo {year} {1984})\BibitemShut {NoStop}%
\bibitem [{\citenamefont {Dresselhaus}(1955)}]{Dresselhaus1955}%
  \BibitemOpen
  \bibfield  {author} {\bibinfo {author} {\bibfnamefont {G.}~\bibnamefont
  {Dresselhaus}},\ }\href@noop {} {\bibfield  {journal} {\bibinfo  {journal}
  {Phys. Rev.}\ }\textbf {\bibinfo {volume} {100}},\ \bibinfo {pages} {580}
  (\bibinfo {year} {1955})}\BibitemShut {NoStop}%
\bibitem [{\citenamefont {D'yakonov}\ and\ \citenamefont
  {Perel'}(1972)}]{Dyakonov1972}%
  \BibitemOpen
  \bibfield  {author} {\bibinfo {author} {\bibfnamefont {M.~I.}\ \bibnamefont
  {D'yakonov}}\ and\ \bibinfo {author} {\bibfnamefont {V.~I.}\ \bibnamefont
  {Perel'}},\ }\href@noop {} {\bibfield  {journal} {\bibinfo  {journal} {Soviet
  Physics Solid State}\ }\textbf {\bibinfo {volume} {13}},\ \bibinfo {pages}
  {3023} (\bibinfo {year} {1972})}\BibitemShut {NoStop}%
\bibitem [{\citenamefont {Lau}\ \emph {et~al.}(2001)\citenamefont {Lau},
  \citenamefont {Olesberg},\ and\ \citenamefont {Flatt\'e}}]{Lau2001}%
  \BibitemOpen
  \bibfield  {author} {\bibinfo {author} {\bibfnamefont {W.~H.}\ \bibnamefont
  {Lau}}, \bibinfo {author} {\bibfnamefont {J.~T.}\ \bibnamefont {Olesberg}}, \
  and\ \bibinfo {author} {\bibfnamefont {M.~E.}\ \bibnamefont {Flatt\'e}},\
  }\href@noop {} {\bibfield  {journal} {\bibinfo  {journal} {\prb}\ }\textbf
  {\bibinfo {volume} {64}},\ \bibinfo {pages} {161301(R)} (\bibinfo {year}
  {2001})}\BibitemShut {NoStop}%
\bibitem [{\citenamefont {Yafet}(1963)}]{Yafet1963}%
  \BibitemOpen
  \bibfield  {author} {\bibinfo {author} {\bibfnamefont {Y.}~\bibnamefont
  {Yafet}},\ }\href@noop {} {\bibfield  {journal} {\bibinfo  {journal} {Solid
  State Physics}\ }\textbf {\bibinfo {volume} {14}},\ \bibinfo {pages} {2}
  (\bibinfo {year} {1963})}\BibitemShut {NoStop}%
\bibitem [{\citenamefont {Li}\ and\ \citenamefont {Dery}(2011)}]{Li2011}%
  \BibitemOpen
  \bibfield  {author} {\bibinfo {author} {\bibfnamefont {P.}~\bibnamefont
  {Li}}\ and\ \bibinfo {author} {\bibfnamefont {H.}~\bibnamefont {Dery}},\
  }\href {\doibase 10.1103/PhysRevLett.107.107203} {\bibfield  {journal}
  {\bibinfo  {journal} {Phys. Rev. Lett.}\ }\textbf {\bibinfo {volume} {107}},\
  \bibinfo {pages} {107203} (\bibinfo {year} {2011})}\BibitemShut {NoStop}%
\bibitem [{\citenamefont {Gmitra}\ \emph {et~al.}(2013)\citenamefont {Gmitra},
  \citenamefont {Kochan},\ and\ \citenamefont {Fabian}}]{Gmitra2013}%
  \BibitemOpen
  \bibfield  {author} {\bibinfo {author} {\bibfnamefont {M.}~\bibnamefont
  {Gmitra}}, \bibinfo {author} {\bibfnamefont {D.}~\bibnamefont {Kochan}}, \
  and\ \bibinfo {author} {\bibfnamefont {J.}~\bibnamefont {Fabian}},\
  }\href@noop {} {\bibfield  {journal} {\bibinfo  {journal} {Phys. Rev. Lett.}\
  }\textbf {\bibinfo {volume} {110}},\ \bibinfo {pages} {246602} (\bibinfo
  {year} {2013})}\BibitemShut {NoStop}%
\bibitem [{\citenamefont {Bir}\ and\ \citenamefont {Pikus}(1974)}]{BirPikus}%
  \BibitemOpen
  \bibfield  {author} {\bibinfo {author} {\bibfnamefont {G.~L.}\ \bibnamefont
  {Bir}}\ and\ \bibinfo {author} {\bibfnamefont {G.~E.}\ \bibnamefont
  {Pikus}},\ }\href@noop {} {\emph {\bibinfo {title} {Symmetry and
  strain-induced effects in semiconductors}}}\ (\bibinfo  {publisher}
  {Halsted},\ \bibinfo {address} {Jerusalem},\ \bibinfo {year}
  {1974})\BibitemShut {NoStop}%
\bibitem [{\citenamefont {Tang}\ \emph {et~al.}(2012)\citenamefont {Tang},
  \citenamefont {Collins},\ and\ \citenamefont {Flatt\'e}}]{Tang2012}%
  \BibitemOpen
  \bibfield  {author} {\bibinfo {author} {\bibfnamefont {J.-M.}\ \bibnamefont
  {Tang}}, \bibinfo {author} {\bibfnamefont {B.~T.}\ \bibnamefont {Collins}}, \
  and\ \bibinfo {author} {\bibfnamefont {M.~E.}\ \bibnamefont {Flatt\'e}},\
  }\href@noop {} {\bibfield  {journal} {\bibinfo  {journal} {Phys. Rev. B}\
  }\textbf {\bibinfo {volume} {85}},\ \bibinfo {pages} {045202} (\bibinfo
  {year} {2012})}\BibitemShut {NoStop}%
\bibitem [{\citenamefont {Restrepo}\ and\ \citenamefont
  {Windl}(2012)}]{Restrepo2012}%
  \BibitemOpen
  \bibfield  {author} {\bibinfo {author} {\bibfnamefont {O.~D.}\ \bibnamefont
  {Restrepo}}\ and\ \bibinfo {author} {\bibfnamefont {W.}~\bibnamefont
  {Windl}},\ }\href {\doibase 10.1103/PhysRevLett.109.166604} {\bibfield
  {journal} {\bibinfo  {journal} {Phys. Rev. Lett.}\ }\textbf {\bibinfo
  {volume} {109}},\ \bibinfo {pages} {166604} (\bibinfo {year}
  {2012})}\BibitemShut {NoStop}%
\bibitem [{\citenamefont {Fedorov}\ \emph {et~al.}(2013)\citenamefont
  {Fedorov}, \citenamefont {Gradhand}, \citenamefont {Ostanin}, \citenamefont
  {Maznichenko}, \citenamefont {Ernst}, \citenamefont {Fabian},\ and\
  \citenamefont {Mertig}}]{Fedorov2013}%
  \BibitemOpen
  \bibfield  {author} {\bibinfo {author} {\bibfnamefont {D.~V.}\ \bibnamefont
  {Fedorov}}, \bibinfo {author} {\bibfnamefont {M.}~\bibnamefont {Gradhand}},
  \bibinfo {author} {\bibfnamefont {S.}~\bibnamefont {Ostanin}}, \bibinfo
  {author} {\bibfnamefont {I.~V.}\ \bibnamefont {Maznichenko}}, \bibinfo
  {author} {\bibfnamefont {A.}~\bibnamefont {Ernst}}, \bibinfo {author}
  {\bibfnamefont {J.}~\bibnamefont {Fabian}}, \ and\ \bibinfo {author}
  {\bibfnamefont {I.}~\bibnamefont {Mertig}},\ }\href@noop {} {\bibfield
  {journal} {\bibinfo  {journal} {Phys. Rev. Lett.}\ }\textbf {\bibinfo
  {volume} {110}},\ \bibinfo {pages} {156602} (\bibinfo {year}
  {2013})}\BibitemShut {NoStop}%
\bibitem [{\citenamefont {Yu}\ and\ \citenamefont {Cardona}(2001)}]{Yu3ed}%
  \BibitemOpen
  \bibfield  {author} {\bibinfo {author} {\bibfnamefont {P.~Y.}\ \bibnamefont
  {Yu}}\ and\ \bibinfo {author} {\bibfnamefont {M.}~\bibnamefont {Cardona}},\
  }\href@noop {} {\emph {\bibinfo {title} {Fundamentals of semiconductors}}},\
  \bibinfo {edition} {3rd}\ ed.\ (\bibinfo  {publisher} {Springer-Verlag},\
  \bibinfo {address} {Berlin},\ \bibinfo {year} {2001})\BibitemShut {NoStop}%
\bibitem [{\citenamefont {Cardona}\ \emph {et~al.}(1988)\citenamefont
  {Cardona}, \citenamefont {Christensen},\ and\ \citenamefont
  {Fasol}}]{Cardona1988}%
  \BibitemOpen
  \bibfield  {author} {\bibinfo {author} {\bibfnamefont {M.}~\bibnamefont
  {Cardona}}, \bibinfo {author} {\bibfnamefont {N.~E.}\ \bibnamefont
  {Christensen}}, \ and\ \bibinfo {author} {\bibfnamefont {G.}~\bibnamefont
  {Fasol}},\ }\href {\doibase 10.1103/PhysRevB.38.1806} {\bibfield  {journal}
  {\bibinfo  {journal} {Phys. Rev. B}\ }\textbf {\bibinfo {volume} {38}},\
  \bibinfo {pages} {1806} (\bibinfo {year} {1988})}\BibitemShut {NoStop}%
\bibitem [{\citenamefont {Jancu}\ \emph {et~al.}(1998)\citenamefont {Jancu},
  \citenamefont {Scholz}, \citenamefont {Beltram},\ and\ \citenamefont
  {Bassani}}]{Jancu1998}%
  \BibitemOpen
  \bibfield  {author} {\bibinfo {author} {\bibfnamefont {J.-M.}\ \bibnamefont
  {Jancu}}, \bibinfo {author} {\bibfnamefont {R.}~\bibnamefont {Scholz}},
  \bibinfo {author} {\bibfnamefont {F.}~\bibnamefont {Beltram}}, \ and\
  \bibinfo {author} {\bibfnamefont {F.}~\bibnamefont {Bassani}},\ }\href
  {\doibase 10.1103/PhysRevB.57.6493} {\bibfield  {journal} {\bibinfo
  {journal} {Phys. Rev. B}\ }\textbf {\bibinfo {volume} {57}},\ \bibinfo
  {pages} {6493} (\bibinfo {year} {1998})}\BibitemShut {NoStop}%
\bibitem [{\citenamefont {Salluzzo}\ \emph {et~al.}(2009)\citenamefont
  {Salluzzo}, \citenamefont {Cezar}, \citenamefont {Brookes}, \citenamefont
  {Bisogni}, \citenamefont {De~Luca}, \citenamefont {Richter}, \citenamefont
  {Thiel}, \citenamefont {Mannhart}, \citenamefont {Huijben}, \citenamefont
  {Brinkman}, \citenamefont {Rijnders},\ and\ \citenamefont
  {Ghiringhelli}}]{Salluzzo2009}%
  \BibitemOpen
  \bibfield  {author} {\bibinfo {author} {\bibfnamefont {M.}~\bibnamefont
  {Salluzzo}}, \bibinfo {author} {\bibfnamefont {J.~C.}\ \bibnamefont {Cezar}},
  \bibinfo {author} {\bibfnamefont {N.~B.}\ \bibnamefont {Brookes}}, \bibinfo
  {author} {\bibfnamefont {V.}~\bibnamefont {Bisogni}}, \bibinfo {author}
  {\bibfnamefont {G.~M.}\ \bibnamefont {De~Luca}}, \bibinfo {author}
  {\bibfnamefont {C.}~\bibnamefont {Richter}}, \bibinfo {author} {\bibfnamefont
  {S.}~\bibnamefont {Thiel}}, \bibinfo {author} {\bibfnamefont
  {J.}~\bibnamefont {Mannhart}}, \bibinfo {author} {\bibfnamefont
  {M.}~\bibnamefont {Huijben}}, \bibinfo {author} {\bibfnamefont
  {A.}~\bibnamefont {Brinkman}}, \bibinfo {author} {\bibfnamefont
  {G.}~\bibnamefont {Rijnders}}, \ and\ \bibinfo {author} {\bibfnamefont
  {G.}~\bibnamefont {Ghiringhelli}},\ }\href@noop {} {\bibfield  {journal}
  {\bibinfo  {journal} {Phys. Rev. Lett.}\ }\textbf {\bibinfo {volume} {102}},\
  \bibinfo {pages} {166804} (\bibinfo {year} {2009})}\BibitemShut {NoStop}%
\bibitem [{\citenamefont {Kahn}\ and\ \citenamefont
  {Leyendecker}(1964)}]{Kahn1964}%
  \BibitemOpen
  \bibfield  {author} {\bibinfo {author} {\bibfnamefont {A.}~\bibnamefont
  {Kahn}}\ and\ \bibinfo {author} {\bibfnamefont {A.}~\bibnamefont
  {Leyendecker}},\ }\href {http://link.aps.org/doi/10.1103/PhysRev.135.A1321}
  {\bibfield  {journal} {\bibinfo  {journal} {Phys. Rev}\ }\textbf {\bibinfo
  {volume} {135}},\ \bibinfo {pages} {A1321} (\bibinfo {year}
  {1964})}\BibitemShut {NoStop}%
\bibitem [{\citenamefont {Mattheiss}(1972)}]{Mattheiss1972a}%
  \BibitemOpen
  \bibfield  {author} {\bibinfo {author} {\bibfnamefont {L.}~\bibnamefont
  {Mattheiss}},\ }\href {http://prb.aps.org/abstract/PRB/v6/i12/p4740\_1}
  {\bibfield  {journal} {\bibinfo  {journal} {Physical Review B}\ }\textbf
  {\bibinfo {volume} {6}},\ \bibinfo {pages} {4740} (\bibinfo {year}
  {1972})}\BibitemShut {NoStop}%
\bibitem [{\citenamefont {Wolfram}(1972)}]{Wolfram1972}%
  \BibitemOpen
  \bibfield  {author} {\bibinfo {author} {\bibfnamefont {T.}~\bibnamefont
  {Wolfram}},\ }\href {\doibase 10.1103/PhysRevLett.29.1383} {\bibfield
  {journal} {\bibinfo  {journal} {Phys. Rev. Lett.}\ }\textbf {\bibinfo
  {volume} {29}},\ \bibinfo {pages} {1383} (\bibinfo {year}
  {1972})}\BibitemShut {NoStop}%
\bibitem [{\citenamefont {Moore}(1949)}]{Moore12}%
  \BibitemOpen
  \bibfield  {author} {\bibinfo {author} {\bibfnamefont {C.~E.}\ \bibnamefont
  {Moore}},\ }\href@noop {} {\emph {\bibinfo {title} {{Atomic Energy Levels. As
  Derived From the Analyses of Optical Spectra}}}},\ Vol.\ \bibinfo {volume}
  {I\&II}\ (\bibinfo  {publisher} {National Bureau of Standards},\ \bibinfo
  {year} {1949})\BibitemShut {NoStop}%
\bibitem [{\citenamefont {van~der Marel}\ \emph {et~al.}(2011)\citenamefont
  {van~der Marel}, \citenamefont {van Mechelen},\ and\ \citenamefont
  {Mazin}}]{Marel2011}%
  \BibitemOpen
  \bibfield  {author} {\bibinfo {author} {\bibfnamefont {D.}~\bibnamefont
  {van~der Marel}}, \bibinfo {author} {\bibfnamefont {J.~L.~M.}\ \bibnamefont
  {van Mechelen}}, \ and\ \bibinfo {author} {\bibfnamefont {I.~I.}\
  \bibnamefont {Mazin}},\ }\href {\doibase 10.1103/PhysRevB.84.205111}
  {\bibfield  {journal} {\bibinfo  {journal} {Phys. Rev. B}\ }\textbf {\bibinfo
  {volume} {84}},\ \bibinfo {pages} {205111} (\bibinfo {year}
  {2011})}\BibitemShut {NoStop}%
\bibitem [{\citenamefont {Lau}\ and\ \citenamefont {Flatt\'e}(2005)}]{Lau2005}%
  \BibitemOpen
  \bibfield  {author} {\bibinfo {author} {\bibfnamefont {W.~H.}\ \bibnamefont
  {Lau}}\ and\ \bibinfo {author} {\bibfnamefont {M.~E.}\ \bibnamefont
  {Flatt\'e}},\ }\href@noop {} {\bibfield  {journal} {\bibinfo  {journal}
  {\prb}\ }\textbf {\bibinfo {volume} {72}},\ \bibinfo {pages} {161311(R)}
  (\bibinfo {year} {2005})}\BibitemShut {NoStop}%
\bibitem [{Note1()}]{Note1}%
  \BibitemOpen
  \bibinfo {note} {Below a temperature of 100K STO undergoes a second-order
  phase transition from cubic to tetragonal structure while oxygens in STO
  start to rotate. \cite {Mattheiss1972a}. This rotation breaks the cubic
  symmetry and causes a further shift in the higher conduction bands, which we
  neglect here.}\BibitemShut {Stop}%
\bibitem [{\citenamefont {Moos}\ \emph {et~al.}(1995)\citenamefont {Moos},
  \citenamefont {Menesklou},\ and\ \citenamefont {H{\"a}rdtl}}]{Moos1995}%
  \BibitemOpen
  \bibfield  {author} {\bibinfo {author} {\bibfnamefont {R.}~\bibnamefont
  {Moos}}, \bibinfo {author} {\bibfnamefont {W.}~\bibnamefont {Menesklou}}, \
  and\ \bibinfo {author} {\bibfnamefont {K.}~\bibnamefont {H{\"a}rdtl}},\
  }\href {\doibase 10.1007/BF01540113} {\bibfield  {journal} {\bibinfo
  {journal} {Applied Physics A}\ }\textbf {\bibinfo {volume} {61}},\ \bibinfo
  {pages} {389} (\bibinfo {year} {1995})}\BibitemShut {NoStop}%
\bibitem [{\citenamefont {Karimov}\ \emph {et~al.}(2003)\citenamefont
  {Karimov}, \citenamefont {John}, \citenamefont {Harley}, \citenamefont {Lau},
  \citenamefont {Flatt\'e}, \citenamefont {Henini},\ and\ \citenamefont
  {Airey}}]{Karimov2003}%
  \BibitemOpen
  \bibfield  {author} {\bibinfo {author} {\bibfnamefont {O.~Z.}\ \bibnamefont
  {Karimov}}, \bibinfo {author} {\bibfnamefont {G.~H.}\ \bibnamefont {John}},
  \bibinfo {author} {\bibfnamefont {R.~T.}\ \bibnamefont {Harley}}, \bibinfo
  {author} {\bibfnamefont {W.~H.}\ \bibnamefont {Lau}}, \bibinfo {author}
  {\bibfnamefont {M.~E.}\ \bibnamefont {Flatt\'e}}, \bibinfo {author}
  {\bibfnamefont {M.}~\bibnamefont {Henini}}, \ and\ \bibinfo {author}
  {\bibfnamefont {R.}~\bibnamefont {Airey}},\ }\href@noop {} {\bibfield
  {journal} {\bibinfo  {journal} {\prl}\ }\textbf {\bibinfo {volume} {91}},\
  \bibinfo {pages} {246601} (\bibinfo {year} {2003})}\BibitemShut {NoStop}%
\bibitem [{\citenamefont {Kalabukhov}\ \emph {et~al.}(2007)\citenamefont
  {Kalabukhov}, \citenamefont {Gunnarsson}, \citenamefont {B\"{o}rjesson},
  \citenamefont {Olsson}, \citenamefont {Claeson},\ and\ \citenamefont
  {Winkler}}]{Kalabukhov2007}%
  \BibitemOpen
  \bibfield  {author} {\bibinfo {author} {\bibfnamefont {A.}~\bibnamefont
  {Kalabukhov}}, \bibinfo {author} {\bibfnamefont {R.}~\bibnamefont
  {Gunnarsson}}, \bibinfo {author} {\bibfnamefont {J.}~\bibnamefont
  {B\"{o}rjesson}}, \bibinfo {author} {\bibfnamefont {E.}~\bibnamefont
  {Olsson}}, \bibinfo {author} {\bibfnamefont {T.}~\bibnamefont {Claeson}}, \
  and\ \bibinfo {author} {\bibfnamefont {D.}~\bibnamefont {Winkler}},\ }\href
  {\doibase 10.1103/PhysRevB.75.121404} {\bibfield  {journal} {\bibinfo
  {journal} {Physical Review B}\ }\textbf {\bibinfo {volume} {75}},\ \bibinfo
  {pages} {121404} (\bibinfo {year} {2007})}\BibitemShut {NoStop}%
\end{thebibliography}%
\end{document}